# Multifunctional Bistable Ultrathin Composite Booms with Flexible Electronics

Yao Yao[1], Juan M. Fernandez[2], Sven G. Bilén[3], and Xin Ning[1,*]

**Abstract**

Small satellites such as CubeSats pose demanding requirements on the weight, size, and multifunctionality of their structures due to extreme constraints on the payload mass and volume. To address this challenge, we introduce a concept of multifunctional deployable space structures for CubeSats based on ultrathin, elastically foldable, and self-deployable bistable composite structures integrated with flexible electronics. The multifunctional bistable booms can be stored in a coiled configuration and self-deploy into a long structure upon initiation by releasing the stored strain energy. The boom demonstrates the capabilities of delivering power and transmitting data from the CubeSat to the flexible devices on the boom tip. The boom also shows the ability to monitor the dynamics and vibration during and after the deployment. A payload boom has been installed in a 3U CubeSat as flight hardware for in-space testing and demonstration. This effort combines morphable ultrathin composite structures with flexible electronics.

**Keywords**

Space Structure, Flexible Electronics, Bistable Structure, Dynamics

[1] Department of Aerospace Engineering, University of Illinois Urbana-Champaign.
[2] Structural Dynamics Branch, NASA Langley Research Center.
[3] School of Engineering Design and Innovation, The Pennsylvania State University.
*Corresponding Author, xning@illinois.edu, 104 S Wright St, 306 Talbot, Urbana, IL 61801

# 1. Introduction

We introduce the concept of an ultrathin, elastically foldable, and self-deployable composite boom structure with integrated flexible electronics. Foldable and deployable space structures that can provide large structural platforms with low storage dimensions are critical components within space systems [1-3]. Advances in deployable space structures have provided solutions for the structural platforms of small satellites, such as CubeSats, for which the volume and mass available for payloads are extremely limited [4, 5]. CubeSats are a class of small satellites consisting of a cube of 10 cm per side (referred to as 1U) or its multiples. The severe constraints on the mass and volume of CubeSats lead to a strong demand for multifunctional structures in which multiple devices and materials can be combined with lightweight and small form-factor structures to serve as both structural platforms and functional payloads. The work presented here aims to address the need for multifunctional deployable space structures for CubeSats.

Bistable booms made from ultrathin carbon fiber–reinforced polymer (CFRP) laminates [6, 7] have been employed as elastically foldable, hinge-free, self-deployable structures to reduce structural mass, size, and deployment complexity [8, 9]. Their thinness (typically below 250 μm) and novel folding and deployment approach offer new opportunities for applications as deployable CubeSat structures. However, they also pose challenges in monitoring their deployment behavior and obtaining multifunctionality.

These ultrathin composite booms often experience large mechanical deformations and complex dynamics during and after deployment, making it difficult to monitor their deployment *in situ*. Non-contact, optical imaging methods using cameras have shown potential in measuring both the static and dynamic deployment of rollable booms without interfering with the mechanical behavior of the boom [10-12]. However, the reliability of the results depends heavily on the camera quality and lighting conditions. Adding additional cameras to mitigate these issues further reduces the feasibility of this approach for small satellites due to the extreme constraints of available space. Contact methods that integrate additional instruments on deployable structures generate more reliable results regardless of the environmental conditions. For example, in a previous effort, a mass was attached to the free end of a bistable boom to help study the internal force generated during self-deployment and compare it to a mathematical model [13]. However, this method changes the structural behavior due to the mass of the attachment. The large deformations and thinness of ultrathin composite booms make it infeasible to integrate traditional thick and bulky electronics for structural monitoring to obtain multifunctionality. Flexible electronics, which are lightweight, thin, and soft, offer a potential solution to these issues [14, 15].

To address these challenges, we developed a novel concept of multifunctional bistable ultrathin composite booms with flexible electronics integrated into a highly compact configuration. The integrated lightweight, thin, and flexible electronics can monitor structural vibration and deployment dynamics of the booms during operation and provide an optical fiducial for imaging purposes. A boom has been integrated into a 3U CubeSat developed by the Virginia Polytechnic Institute and State University (Virginia Tech) that is tentatively scheduled for launch circa 2025. This paper presents the design and development, as well as fundamental studies of the mechanics and dynamics of a multifunctional bistable boom.

The paper is organized as follows. Section 2 presents the design, materials, and fabrication methods to realize the multifunctional booms integrated with flexible electronics. Section 3 includes the results and discussions on the space environmental tests, on-ground vibration tests, and deployment tests. Section 4 concludes the paper.

## 2. Material and Methods

### 2.1 Rollable Bistable Boom with Flexible Electronics

The payload boom consisted of a bistable ultrathin CFRP composite shell, flexible electronic devices on the boom tip, and thin conductive wires on the surface embedded within the shell (Fig. 1). Compared to conventional composite laminates with typical thickness in the order of millimeters to centimeters, the total thickness of this ultrathin composite laminate with three plies was 156 μm. An ultrathin thickness was required to enable elastic folding and to minimize the volume of the coiled structure when stored in the CubeSat. Any functional devices integrated with the booms must accommodate the aggressive deformation of the booms during coiling and deployment without damage to themselves or the host boom structures. The functional devices were required to sense the translational and rotational motions of the boom tip during deployment and post-deployment vibrations. Additionally, a light-emitting diode (LED) was required at the boom tip to serve as an optical fiducial for an on-board camera to provide visual confirmation of the boom deployment. The boom needed to pass power to the integrated electronics and transmit data back to the CubeSat flight computer. All electronics must operate and survive in the harsh thermal–vacuum conditions of space.

To meet these requirements, we developed a deployable structure for CubeSats based on the concept of a bistable ultrathin composite boom (Fig. 1a and b). The boom was a tape-measure-like shell with a parabolic cross-section (Fig. 1c). It was 1.22 m long in a fully extended configuration, and approximately 70 mm wide in a coiled state. It was stable in its fully coiled configuration and achieved self-deployment to a fully extended shape by releasing its stored strain energy. A bistable boom design was employed by NASA Langley Research Center (LaRC) to simplify the CubeSat mechanisms for storing and deploying the structure. The flexible devices included a six-axis inertial measurement unit (IMU), an LED, other supporting electronic components, and circuitry, which were attached to the surface of the boom tip (Fig. 1d).

Power and data transmission wires were required to run along the entire boom length to connect the boom-tip flexible devices to the CubeSat data acquisition (DAQ) and power distribution units. The maximum thickness of the embedded wires needs to be approximately 50 μm to minimize the mass and thickness added to the structure. Moreover, a practical requirement was that the wires needed to be readily available with high quality and high consistency in electrical characteristics for replicable and predictable performance. Therefore, we selected commercially available 44-AWG wires over other technologies initially investigated for this application, such as printed conductive ink traces. A 44-AWG wire has a diameter of 50.2 μm, including a layer of enamel coating for electrical insulation. We fabricated two multifunctional boom designs with either flexible thin wires for power and data transmission embedded within the laminate or attached to the boom surface (Fig. 1e). The boom design with embedded transmission wires was selected as the flight hardware installed on the 3U CubeSat, whereas the one with surface wires was used for ground-test experiments.

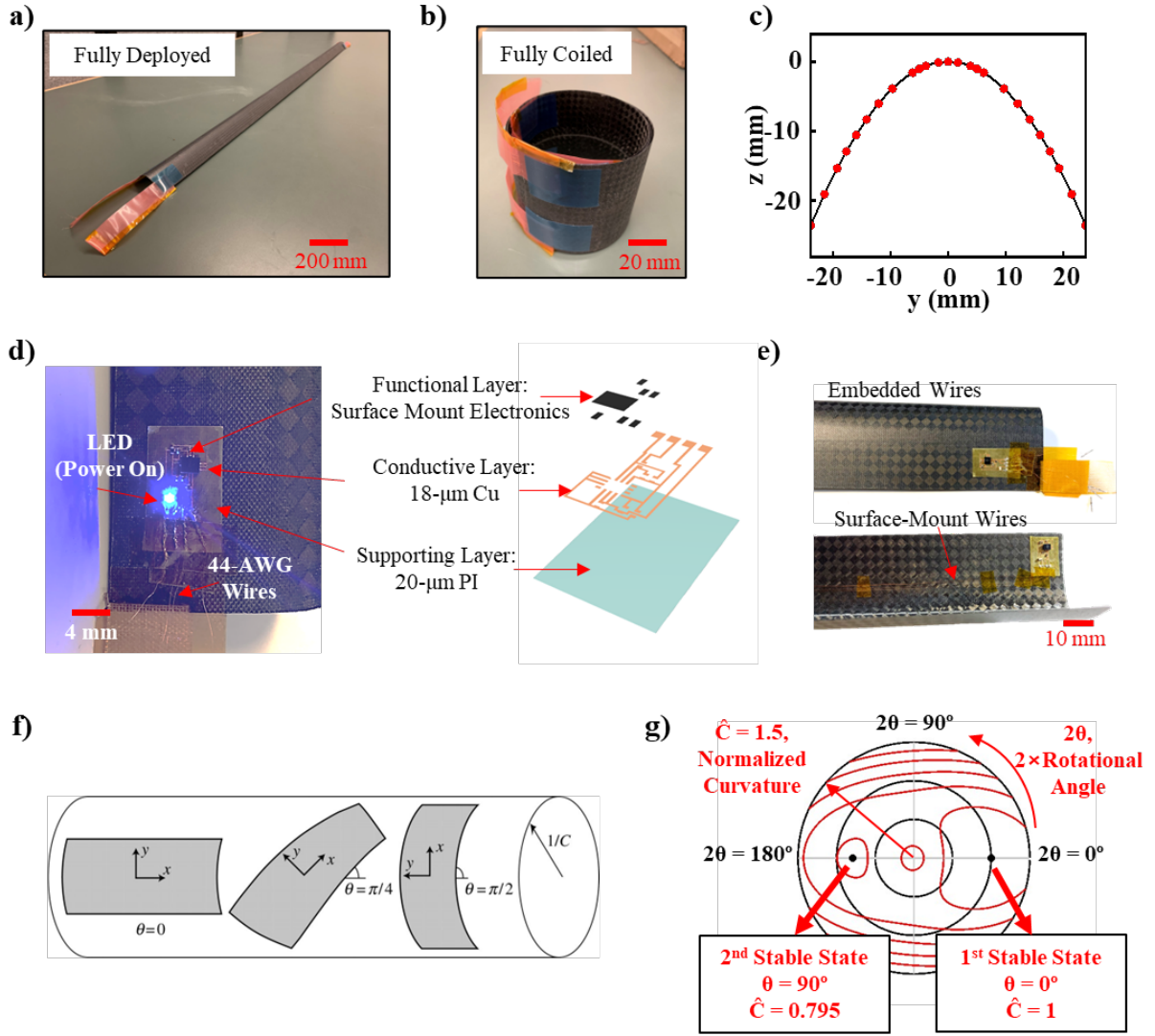

**Fig. 1 Bistable ultrathin composite boom integrated with flexible electronics.** a) The boom in the fully extended state. b) The boom in its fully coiled state. c) Boom cross-section with reference points for shape constructions in modeling. d) Optical image and exploded schematic of the boom-tip flexible devices. e) Boom-tip flexible devices connected to the embedded or surface-mount 44-AWG wires. f) Coordinate system used to define the shell configuration [16]. g) Polar contour plot of normalized bending strain energy as a function of the non-dimensional curvature and the shell angle (two stable local minima points marked as dots show the two stable states).

The flight-hardware boom installed on the CubeSat included a total of 16 wires embedded inside the laminate stack. The embedded wires were divided into two groups of 8 wires symmetrically distributed on each end of the parabolic cross-section. Such design was beneficial for protecting the long-span wires from the space environment and limiting the total boom thickness to below 200 μm. The wires were manually laid up under tension to keep them straight during the construction of the boom laminate. The wires extended 12 cm past each end of the booms to subsequently enable electrical connections to the other components. As for the boom for ground testing and dynamics studies, the wires were attached loosely to the boom surface using Kapton tape at a few locations. Since the attached wires were not under tension, they could not provide additional mechanical forces to alter the structural behavior. Additionally, the total mass of the attached wires and the tape used was approximately 0.2 g, which was less than 1% of the total mass of the boom (20.4 g). We used the boom with surface-mount wires for studying the deployment dynamics due to the limited number and high manufacturing costs of the flight-hardware booms with embedded wires.

## 2.2 Materials and Mechanics of the Bistable Ultrathin Composite Boom

The 1.22-m-long boom in this study was a bistable ultrathin tape measure–like structure with a parabolic cross-section designed by NASA LaRC [17]. Fig. 1d shows the parabolic boom cross-section, where the discretization points were from an analytical model used to generate the parabolic geometry [17]. The boom was fabricated at NASA LaRC from a symmetric composite laminate consisting of three plies. Two carbon fiber–epoxy (M30S/PMT-F7[4]) plain weave (PW) plies served as the top and bottom layers of the stack, which were oriented at ±45° to the boom axial or longitudinal direction in the deployed configuration. A single carbon fiber–epoxy (MR60H/PMT-F7) unidirectional (UD) ply at 0° (along the longitudinal axis) was sandwiched between the two PW layers.

The bistability effect was generated by a combined effect of composite laminate layup and geometry [18]. This bistable boom had a first stable state with zero strain energy, which was the as-manufactured extended configuration (Fig. 1a), and a second stable state with a higher strain energy at the coiled configuration. The coiled boom did not need any constraints to keep its coiled configuration. Given the boom geometry and laminate properties, an analytical model was used to predict the strain energy and diameter of the coiled configuration of the boom [17]. The model was described in detail previously by Lee et al. [17] with the key theoretical method summarized here.

The mid-plane of a cylindrical shell was assumed to only bend without stretching under plane stress, and the Gaussian curvature remains identical to its stress-free configuration. The state of the shell configuration can be described by the cylinder's curvature $C$ and the shell's rotation relative to the cylinder's longitudinal axis, $\theta$ (Fig. 1f) [16].

In this study, due to the non-circular cross-section of the boom, the cross-section was approximated by discretized circular segments that have different radius of curvature $R_i$, subtended angle $\alpha_i$, and arc length $L_i = R_i \alpha_i$ (subscript $i$ denotes each arc segment). In addition, to maintain the continuity of the geometry, the segments shared the same tangent at the start and end points with their neighboring segments. The non-dimensional change in the shell curvature $\widehat{\boldsymbol{k}_\iota}$ of the $i$-th segment can then be calculated as [19]:

$$\widehat{\boldsymbol{k}_\iota} = R_\text{e} \Delta \begin{bmatrix} k_x \\ k_y \\ k_{xy} \end{bmatrix} = \frac{\hat{C}}{2} \begin{bmatrix} 1 - \cos 2\theta \\ \cos 2\theta + 1 - \frac{2}{\hat{C}}\left(\frac{R_\text{e}}{R_i}\right) \\ 2\sin 2\theta \end{bmatrix}, \qquad (1)$$

where $R_\text{e}$ was the normalizing radius, which ensured that the non-dimensional curvature $\hat{C} = C R_\text{e} = 1$ when the shell was placed at its initial extended state.

Then, the bending strain energy per unit length $\widehat{U}$, which was the total strain energy of all the circular shell segments, can be calculated as [20]:

$$\widehat{U} = \frac{1}{2L_\text{p}} \sum_{i=1}^{n} L_i \widehat{\boldsymbol{k}_\iota}^T \widehat{\boldsymbol{D}} \widehat{\boldsymbol{k}_\iota}, \qquad (2)$$

---

[4] Specific vendor and manufacturer names are explicitly mentioned only to accurately describe the test hardware. The use of vendor and manufacturer names does not imply an endorsement by the U.S. Government, nor does it imply that the specified equipment is the best available.

where $L_\text{p}$ was the total length of the parabolic cross-section, and $\widehat{\boldsymbol{D}}$ was a non-dimensional bending stiffness matrix that was calculated using the classical laminate theory [21]. The Appendix includes the calculation and values of the stiffness matrix of the composite materials.

The stable state configuration was found by minimizing the bending strain energy:

$$\delta \widehat{U} = \frac{\partial \widehat{U}}{\partial \theta} \delta\theta + \frac{\partial \widehat{U}}{\partial \hat{C}} \delta \hat{C} = 0, \tag{3}$$

where $\theta$ and $\hat{C}$ satisfied:

$$\frac{\partial \widehat{U}}{\partial \theta} = 0, \quad \frac{\partial \widehat{U}}{\partial \hat{C}} = 0. \tag{4}$$

To incorporate the effect of the embedded copper wires, the longitudinal modulus of the mid-plane ply was modified using a homogenization method based on the volume fraction of the embedded copper wires. The modified longitudinal modulus of the mid-plane ply, $E_{11}$, was calculated to be 142.6 GPa, whereas, for comparison, the unmodified $E_{11}$ was 144.1 GPa. The ABD matrix (Appendix, Eq. A7) of the composite shell with embedded wires was then calculated by using the modified $E_{11}$. The effects of embedded wires on the transverse modulus were negligible.

Fig. 1g presents a contour plot of the energy landscape of the boom in polar coordinates. The angular coordinate was denoted by $2\theta$, which was two times the shell's relative orientation to the cylinder's longitudinal axis, $\theta$, shown in Fig. 1f. The radial coordinate was denoted by $\hat{C}$, which was the non-dimensional curvature of the shell, defined in Eq. 1. Two local minima points were marked in Fig. 1g and correspond to the two stable states of the bistable boom. The analytical model predicted the coiled boom without embedded wires had a diameter of 78.3 mm (or a non-dimensional curvature $\hat{C}$ = 0.7948). The boom with embedded wires retained two stable states at $\theta$ = 0° (deployed shape) and $\theta$ = 90° (coiled shape). Compared to the boom without embedded wires, the non-dimensional curvature $\hat{C}$ remained the same for the deployed state ($\hat{C}$ = 1) and showed negligible changes for the coiled state ($\hat{C}$ = 0.7952) with only 0.05% difference. Therefore, the effects of embedded wires on the geometries of deployed and coiled booms can be ignored.

**2.3 Boom-tip Flexible Devices**

The flexible devices included an IMU motion sensor, supporting electronic components, and circuitry that were attached to the surface of the boom tip. Fig. 1d shows an optical image and an exploded schematic of the circuitry of the boom-tip flexible devices. The devices had three layers of material: a 20-µm-thick polyimide layer as the substrate material, 18-µm-thick copper traces as the conductive material to form the circuit interconnections, and a functional layer containing all the surface-mount resistors, capacitors, and MEMS (micro-electromechanical system)-based IMU. Copper was used to form the circuit interconnects due to its excellent electrical conductivity and low cost. The surface-mount IMU (LSM6DSO32, STMicroelectronics) used in the current design was a six-axis inertial module, which contained a three-dimensional (3D) digital accelerometer and a 3D digital gyroscope. This IMU had an operating temperature range from −40 to 85 °C, suitable for space applications. The accelerometer and gyroscope had a noise density of 120

μg/$\sqrt{\text{Hz}}$ and 3.8 mdps/$\sqrt{\text{Hz}}$. An LED (Würth Elektronik, 150060BS75000) was added between the power lines as an indicator of successful power transfer as well as of the location of the boom-tip flexible devices while operating in space.

Microfabrication was used to fabricate the flexible circuits. Solder bonding was then employed to integrate the commercial-off-the-shelf (COTS) components. The fabrication of the flexible devices started with attaching the polyimide–copper laminate (AR1820, Dupont) to a 75 mm × 50 mm glass slide with an adhesive layer of polydimethylsiloxane (PDMS) spun-coated at 1000 rpm for 60 seconds. The thin-film PDMS was partially cured at 110 °C for 90 seconds so that the attachment was firm but could be released after microfabrication. The detailed circuit design is presented in Fig. S2 (Appendix). To define the pattern on the polyimide–copper laminate, a layer of AZ 5214 Image Reversal Photoresist was spin-coated on the laminate surface at 3000 rpm for 30 seconds. Photolithography defined the photoresist with the designed pattern. To remove the exposed copper, the laminate was immersed in Copper Etchant type 100 for 20 seconds and repeatedly rinsed with de-ionized (DI) water for 10 seconds until the bottom polyimide layer was fully exposed. The fabricated circuit was then rinsed with acetone, isopropanol, and DI water. After the laminate was released from the glass slide, all electronic components were soldered using low-temperature solder paste (TS391LT, Chip Quick Inc.) and a hot-air rework station. The total mass of the fabricated flexible device was approximately 0.05 grams, which was a 0.25% added mass on the boom tip, compared with the total mass of the bistable composite boom (20.4 grams). The boom-tip flexible devices were attached to the boom surface by Kapton tape and were electrically connected through the 44-AWG magnet wires by direct soldering. The power consumption of the boom-tip flexible devices was measured. The surface-mount IMU used 0.7 mW at 3.4 V on average, whereas the LED used 86.4 mW when lit at 3.4 V.

### 2.4 DAQ and Communication Protocol

The boom was required to be wrapped around a rotating spool in the CubeSat. The root of the boom was mechanically clamped and electrically connected to the spool integrated with a proximity board (Fig. S1) that stored data during deployment and retraction. The proximity board used a slip ring to transmit data to a stationary data acquisition (DAQ) system and received power from the CubeSat. The flexible devices attached at the boom tip were electrically connected to the proximity board through the thin conductive wires.

Inter-Integrated Circuit (I$^2$C) protocol was used for the communication between the boom-tip electronics and the DAQ. The I$^2$C protocol can support communications between multiple controller units (i.e., the DAQ) and multiple peripheral units (i.e., the boom-tip flexible devices) with only four lines, regardless of the number of connected units. As shown in Fig. S2, the power transfer lines included the supply voltage (VDD) and ground (GND), whereas the data transfer lines included the serial clock (SCL) and serial data (SDA). Therefore, only four embedded wires were required even if multiple devices were used. The IMU used in this study had six 16-bit registers storing the measured data from both the 3-axis gyroscope and 3-axis accelerometer and a 32-bit register storing the timestamp of each sample. With the current setup, each of the data registers took 0.5 ms to complete data transmission (1 ms for the timestamp register), resulting in a total of 4 ms for the data transmission time. This was sufficient in supporting the 208-Hz sampling rate of the IMU.

### 3. Results and Discussion

## 3.1 Stability of Boom-tip Devices in Thermal–Vacuum Conditions

To prove that the fabricated devices are suitable for space application, we assessed the survivability and stability of these units in a simulated space environment. The flexible devices were placed inside a thermal–vacuum chamber (TVAC) that simulates the thermal conditions and high vacuum of low-Earth orbit (LEO). Fig. 2a presents a schematic and a picture of the experimental setup. The boom-tip flexible devices were attached with polyimide tape on a thermal plate inside the TVAC unit. An Arduino Uno (DAQ) outside the chamber connects the flexible devices and the 44-AWG wires through a DB9 connector. The baseline response at rest for all six axes of the IMU under normal room conditions is shown in Fig. 2b. Since only the $z$-axis of the accelerometer was aligned with gravity, it was the only axis showing a reading of 1 g. All readings from the other axes remained zero. Similar data acquired at two other temperatures (7 °C and 50 °C) and in vacuum (approximately at $1 \times 10^{-6}$ Torr) are presented in Fig. 2c and 2d. Additional data at other temperatures between 7 °C to 50 °C were acquired and showed similar results.

The baseline noise for all axes increased due to the vibration of the chamber caused by the vacuum pumps and the TVAC's cooling and heating system. However, the averaged readings from all axes exhibited no change. The temperature was measured with an internal temperature sensor on the boom-tip IMU. Although the actual thermal plate temperature was set to a wider range (−40 °C to 60 °C), it was discovered that the internal temperature of the boom-tip IMU is different from the external temperature. This was due to the internal heating of the boom-tip IMU while it was powered and a limited thermal conduction area between the thermal plate and IMU under vacuum.

## 3.2 Vibration of Deployed Multifunctional Boom

The integrated flexible devices were used to measure the vibration of the deployed bistable boom to study the vibrational modes and validate the device functions. Two different testing methods were employed: impact-induced vibration and continuously excited vibration. The results of each method are presented in Sections 3.2.1 and 3.2.2. Finite element analysis (FEA) results are in Section 3.2.3. Comparison of the results across different testing setups as well as the FEA results are presented in Section 3.2.4.

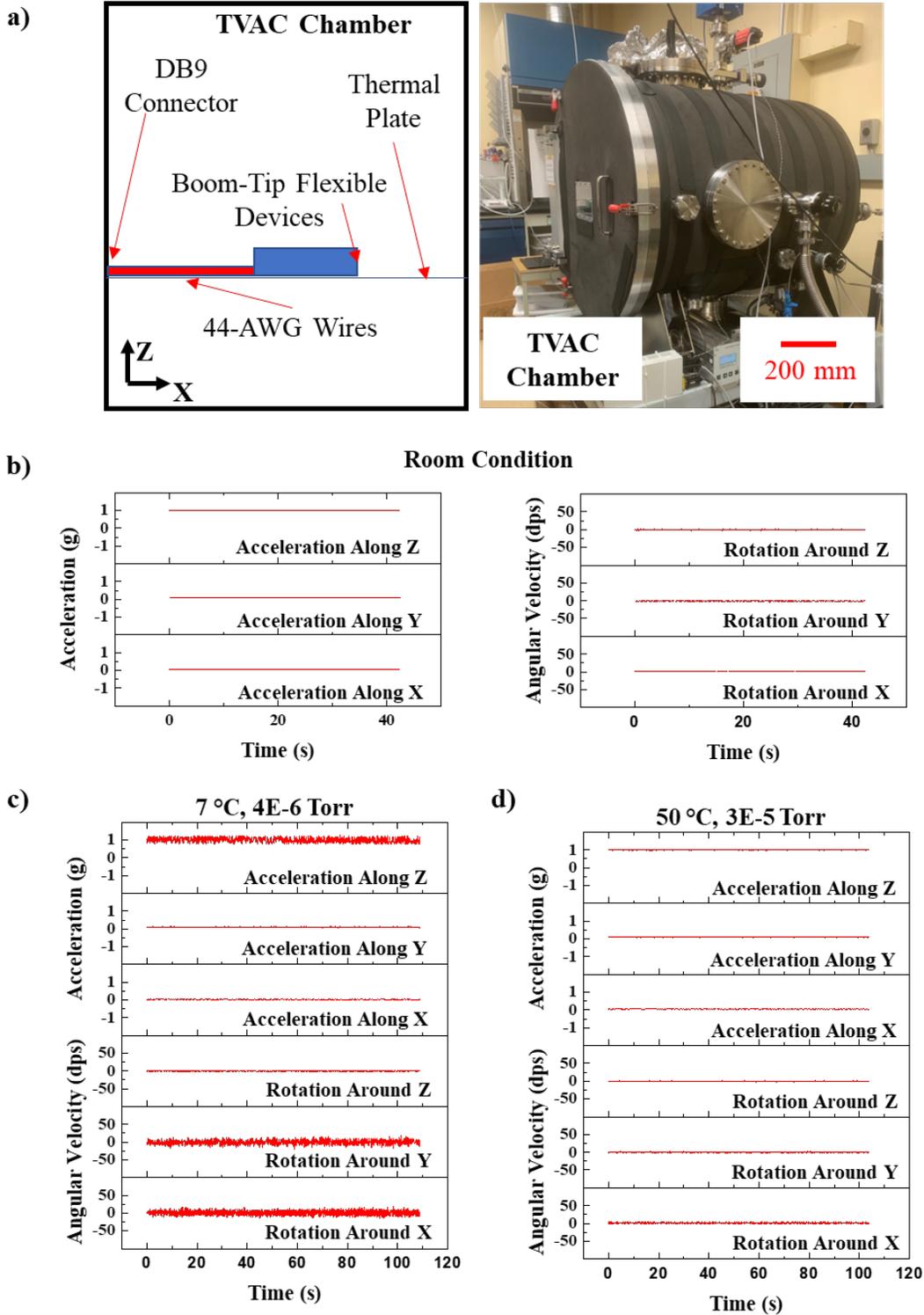

**Fig. 2 Setup and Results of the Survivability and Stability Experiments.** a) Schematic and optical image of the experimental setup of the thermal–vacuum tests. b) Baseline reading of all 6 axes of the boom-tip IMU at: b) room condition, c) 7 °C, 4×10⁻⁶ Torr, and d) 50 °C, 3×10⁻⁶ Torr.

### 3.2.1 Impact-Induced Vibration

Fig. 3a shows the experimental setup for the impact-induced vibration test. The boom was fixed to a 3D-printed clamp, which has two parts to fit with both the concave and convex surfaces of the boom. The cross-section of the clamp surface, interacting with the bistable boom surface, was identical to that of the boom shown in Fig. 1c. The clamp parts were held together by two stainless-steel bolts on both ends of the parabola. The length of the clamp along the boom axial direction was 60 mm. To excite vibrations in the bistable boom, an impulse was manually generated by displacing one corner of the boom tip and quickly releasing it. The boom was allowed to vibrate until the vibrations were naturally damped. Note that the shaker shown in Fig. 3a was not used for impact-induced vibration but was employed in the next section.

Angular rotational velocity and linear acceleration were recorded along all three axes of the IMU. The $x$-, $y$-, and $z$-axes denote the longitudinal, transverse, and surface normal directions of the boom, respectively. A sampling frequency of approximately 200 Hz was used. The range of the IMU measurements was set to ±1000 degrees per second (dps) for angular velocity and ±8 g for linear acceleration.

Figures 3b and c present the time-domain signals of angular velocity and linear acceleration for the boom at its full length extended from the clamp (1162.2 mm). The tests demonstrated that the IMU integrated on the boom tip successfully captured the dynamic processes. The measurements showed that the rotational and translational vibrations damped out within 10 to 15 seconds. Rotation around the $x$-axis, i.e., longitudinal direction, exhibited significantly larger amplitude than rotations in other directions (Fig. 3b), indicating that the twisting around the $x$-axis was a dominating motion. The results showed that maximum accelerations in $y$- and $z$-axes were five times larger than the one along the $x$-axis. Acceleration in the $y$-axis (i.e., boom transverse direction) was generated mainly due to the horizontal motion (parallel to the boom surface). The $z$-direction acceleration was from the $z$-direction (perpendicular to the boom surface) translational motion (Fig. 3c).

To further analyze the vibrations, we used a Fourier transform to obtain the power spectral density (PSD) in the frequency domain (Fig. 3d and 3e). The analyses utilized a Hanning window containing 2000 data points and 1000 overlapping points that provided a frequency resolution of approximately 0.1 Hz. Several peaks were identified in the PSD plot. We can identify the expected vibration signals for lower-order vibration modes based on the PSD. The first major vibration mode occurred at 5.2 Hz. Since the largest amplitude was observed in the angular velocity around the $x$-axis, it was determined that this mode was a twisting motion around the boom longitudinal direction, i.e., the first twisting mode. The second peak identified was at 19.4 Hz, where, again, the angular velocity around the $x$-axis dominated the motion, which indicates the second twisting mode around the longitudinal axis. Due to the nature of the impact, where one corner of the boom tip was deflected, only twist modes were primarily excited. Therefore, to excite the bending mode of the boom, other excitations were performed.

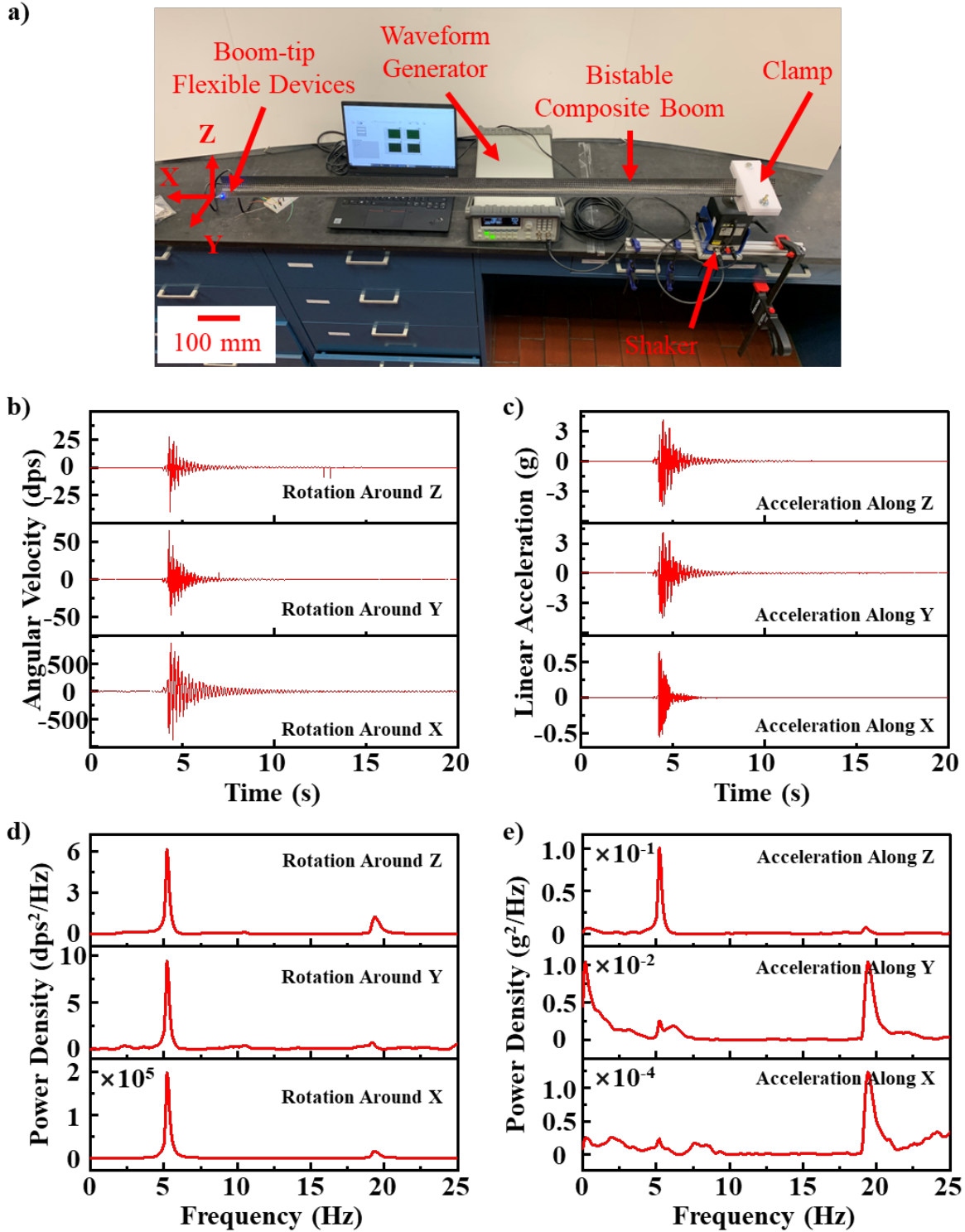

**Fig. 3 Experimental setup and results of impact-induced vibration.** a) Image of a 1.22-m bistable composite boom mounted on a table with a 3D-printed clamp. The boom-tip flexible devices are attached to the boom surface with Kapton double-sided tape and are electrically connected to the DAQ through surface-mount 44-AWG wires. The axes are labeled and used for further analysis. The $x$-, $y$-, and $z$-axes point along the longitudinal, transverse, and normal directions of the boom. b) and c) Time-domain rotation and acceleration at the boom tip during impact-induced vibration. d) and e) Power spectral density plots of the impact-induced vibration.

**3.2.2 Continuously Excited Vibration**

The impact-induced vibrations damped out quickly so that the modes with smaller amplitudes diminished without being captured in the PSD results. To further extend the duration of the excited vibration modes, a shaker (Mini SmartShaker™ model K2007E01) was used to continuously excite the system. Fig. 3a shows the experimental setup for monitoring the continuously excited vibration. The clamp was connected to the actuator of the shaker through a stainless-steel rod. The asymmetry of the placement of the shaker mounting piece was intentional to excite vibrations in both twisting and bending directions. The sampling parameters of the IMU and the DAQ systems remained identical to those provided in Section 3.2.1.

The shaker was driven by a waveform generator to provide the desired actuation output. In this study, we used two different waveforms to excite the shaker: white noise and linear chirp spectrum. The white noise signal was generated by the waveform generator with a root-mean-square voltage (Vrms) of 1 V. With the white noise signal, the data acquisition period was prolonged to yield more accurate PSD results. For comparison, a linear chirp spectrum was generated with a starting frequency of 1 µHz and a frequency increase rate of 0.5 Hz/s. The PSD results of the white-noise-induced vibrations and the linear-sweep-induced vibrations are shown in Fig. 4. Similarly, the first twisting mode was observed at 5.2 Hz with white-noise excitation and 5.1 Hz with linear-sweep excitation. The second twisting mode was observed at 19.0 Hz with white-noise excitation and at 19.5 Hz with linear-sweep excitation. Additionally, the first bending mode was successfully excited. A peak at 12.7 Hz with white-noise excitation and at 12.1 Hz with linear-sweep excitation were determined to be the first bending mode. The results were further confirmed with finite element analysis (FEA) simulations.

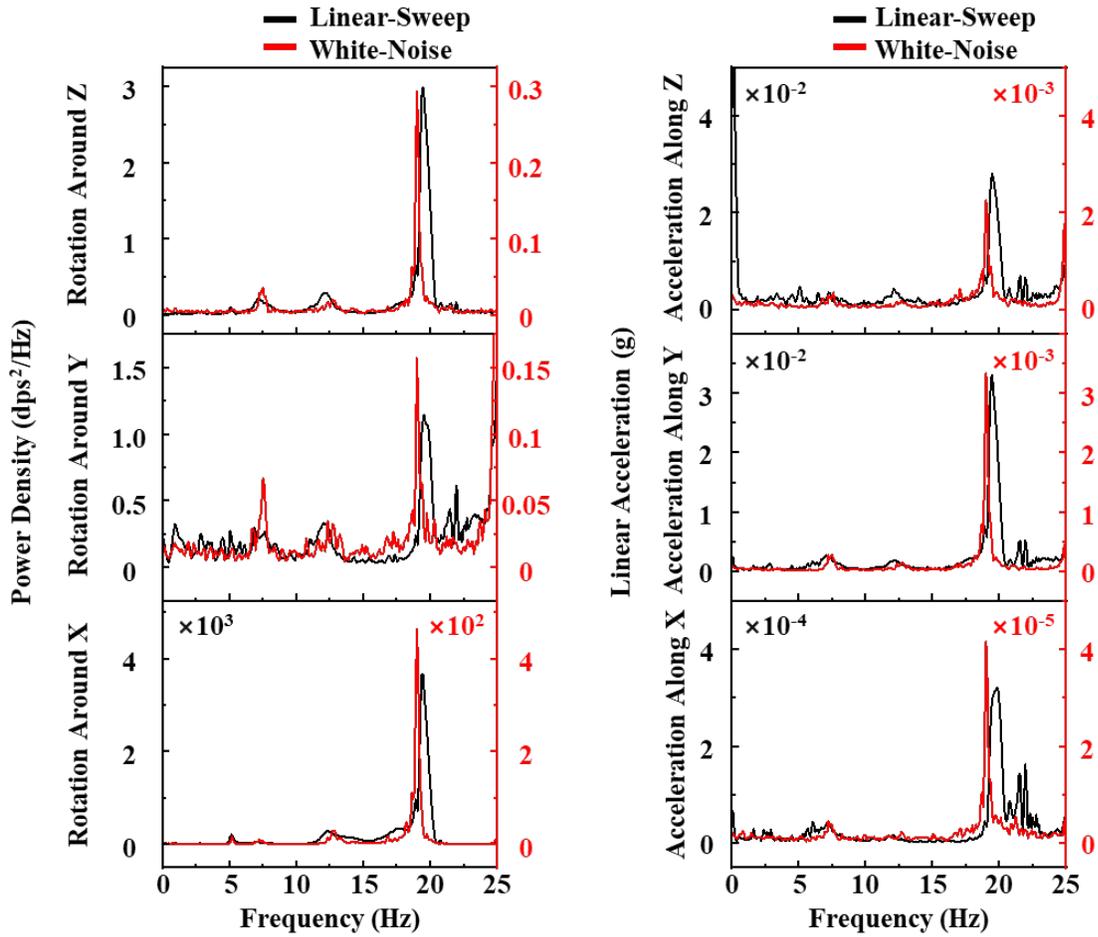

**Fig. 4 Power spectral density plots for monitoring continuously excited vibration.**

### 3.2.3 Finite Element Analyses

FEA simulations using Abaqus provided the numerical results of vibration modes and frequencies of the boom. The cross-section was defined with spline according to the dimensions shown in Fig. 1d. The simulated boom part was defined as a thin shell structure with a composite layup ([45PWc/0c/45PWc]) and density (1552.4 kg/m$^3$). The composite material properties are shown in Table 1.

**Table 1 Material properties of thin-ply composites for FEA simulation.[22]**

| Label | Material Form | $E_1$ (GPa) | $E_2$ (GPa) | $\nu_{12}$ | $G_{12}$ (GPa) | Thickness $t$ (μm) |
|---|---|---|---|---|---|---|
| c | Unidirectional Carbon Fiber | 144.1 | 5.2 | 0.335 | 2.8 | 40.0 |
| PWc | Plain Weave Carbon Fiber | 89.0 | 89.0 | 0.035 | 4.2 | 58.2 |

The boom had a uniform mesh size of 2.3 mm and was defined with S4R shell elements. A clamp located near the end of the boom was modeled as an elastic solid polylactide (PLA) material, with a density of 1100 kg/m$^3$ and Young's modulus of 4.2 GPa, to simulate the actual clamping condition. The bottom and top surfaces of the PLA clamp were fixed in space by allowing zero displacement and rotation. A linear perturbation method (Lanczos) served as the solver to calculate the vibrational modes and frequencies.

Figure 5 presents the shape and frequencies of the first three vibrational modes, and Supplemental Videos 1–3 show the vibration. The lowest-frequency mode at 5.31 Hz was a twisting motion around the *x*-axis, followed by a bending mode with translational motion in the vertical direction (*y-z* plane) at 14.08 Hz. Furthermore, there existed a second twisting mode at 20.83 Hz.

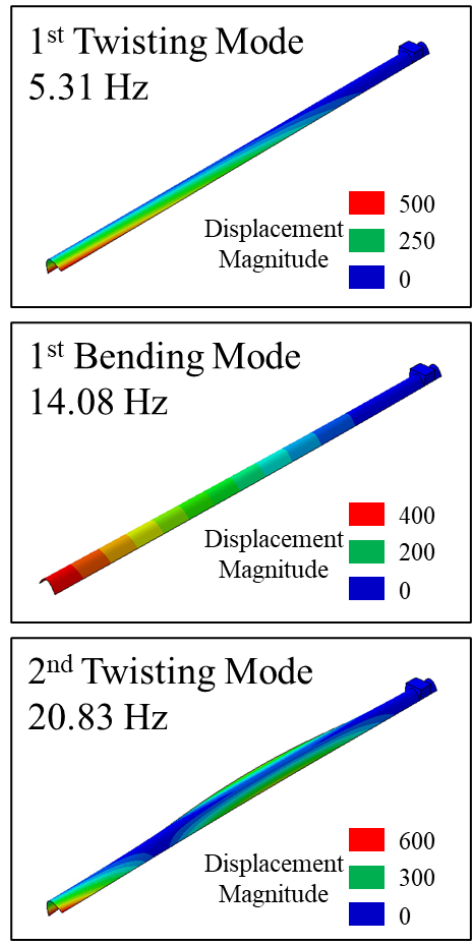

**Fig. 5 FEA frequency analysis results.** The first three vibration frequencies and corresponding modes are shown.

### 3.2.4 Discussions on Experimental and Numerical Results

Table 2 includes all the natural frequencies from the vibration tests and FEA. The experimental results obtained through various excitation methods agreed well with each other within a 5% difference. The differences between experimental frequencies and simulation results were between 5 to 15%. There might have been two reasons leading to the discrepancies. First, material defects during the fabrication process could not be accurately considered in the FEA simulations, which might have led to reduced stiffness and, therefore, lower frequencies in the experiments. Second, the measured frequencies included the effects of the entire testing system such as shaker, clamps, test table, etc., which resulted in weaker boundary conditions and lower natural frequencies.

**Table 2 Experimental and FEA results of natural frequencies and corresponding modes.**

| Vibration Mode | Natural Frequencies (Hz) | | | |
|---|---|---|---|---|
| | Simulation | Impact | White Noise | Linear Sweep |
| Mode 1 (1st Twisting) | 5.31 | 5.2 | 5.2 | 5.1 |
| Mode 2 (1st Bending) | 14.08 | N/A | 12.7 | 12.1 |
| Mode 3 (2nd Twisting) | 20.83 | 19.4 | 19.0 | 19.5 |

### 3.3 Tip Roll-out Deployment

We conducted a boom roll-out self-deployment test to validate the effectiveness of boom-tip devices in monitoring the fast dynamic process and to study the dynamics and post-deployment vibration. Figure 6a shows the testing process, and Supplemental Video 4 is the recorded video. The tip of the boom with the attached flexible devices was coiled up to store strain energy for deployment. After approximately four cycles of rotation, the boom was fully coiled and then held steady. Upon release, the boom self-deployed by releasing the stored strain energy. The self-deployment took approximately 1 second and Supplemental Video 4 shows a clear twisting vibration after the full deployment.

Figures 6b and 6c present the acceleration and angular velocity recorded by the boom-tip devices. During deployment, the motion amplitude in both linear acceleration and angular velocity exceeded the maximum data acquisition range of the IMU: ±32 g for linear acceleration and ±2000 dps for angular velocity. Therefore, this roll-out deployment approach experienced dynamics that require IMUs with larger measurement ranges. It also indicated that this deployment mechanism is an aggressive process that may not be desirable in practical applications. However, it should be noted that the flexible devices on the boom tip survived the coiling and dynamic deployment processes.

PSD results were calculated and plotted for the signal after the full deployment, thereby revealing the post-deployment frequencies (Fig. 6d and 6e). The first frequency captured was at 5.31 Hz, which corresponded to the 1st vibration mode (i.e., 1st twisting mode) of the boom. It could also be confirmed in Table 2 through the FEA. Another mode at 14.07 Hz was observed in the *z*-axis acceleration, corresponding to the 1st bending mode. Due to the absence of the shaker in this test, the experimental clamping condition was stronger and closer to the simulated boundary condition. Therefore, the measured vibrational frequencies matched better with the simulated vibrational frequencies with less than 1% difference.

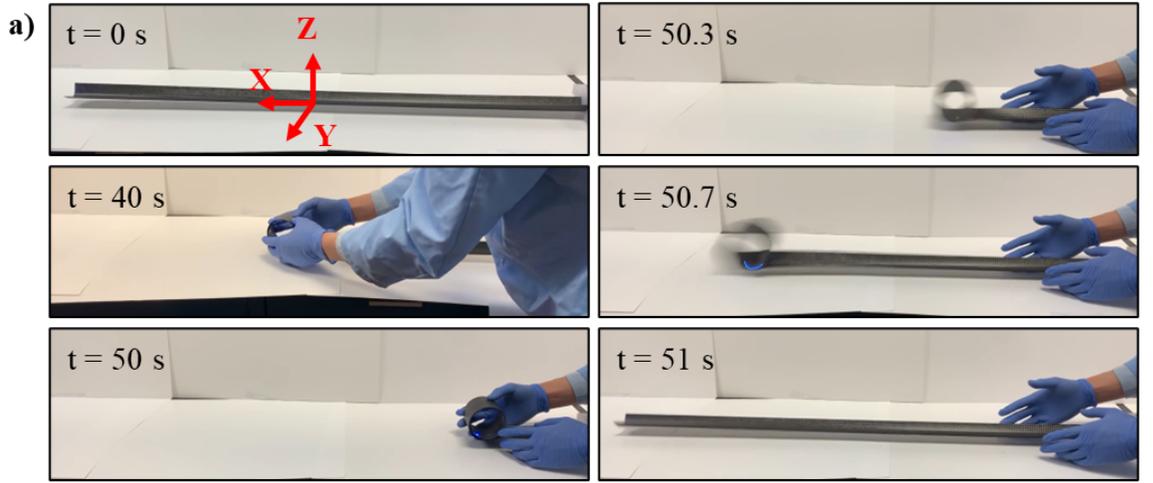
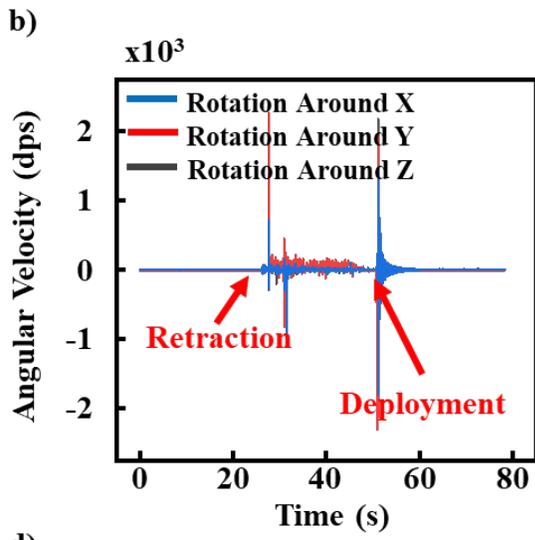
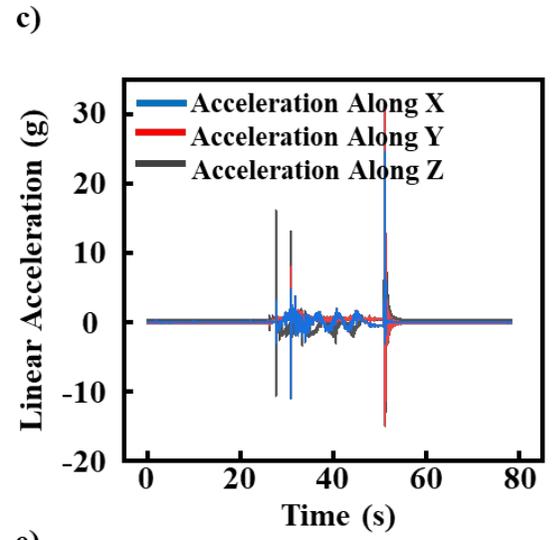
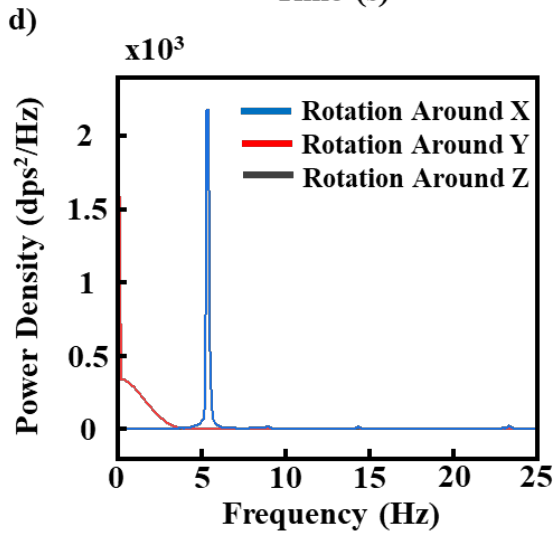
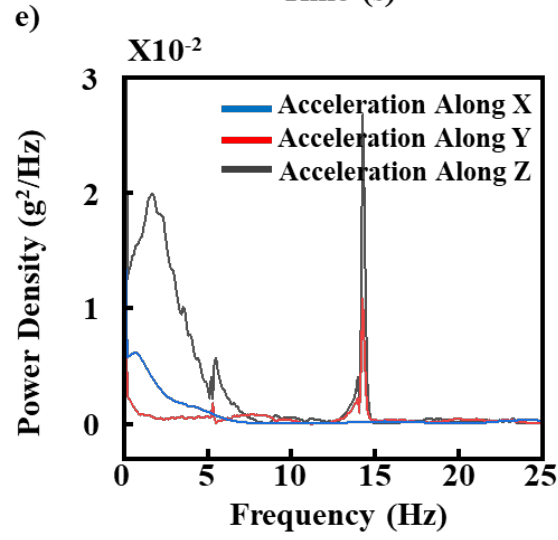

**Fig. 6 Roll-out Deployment Test and Vibration Analyses.** a) The coiling and self-deployment process. b) and c) Time-domain data of rotation and acceleration during the coiling and self-deployment process. d) and e) Power spectral density plots of the dynamic behavior after full deployment.

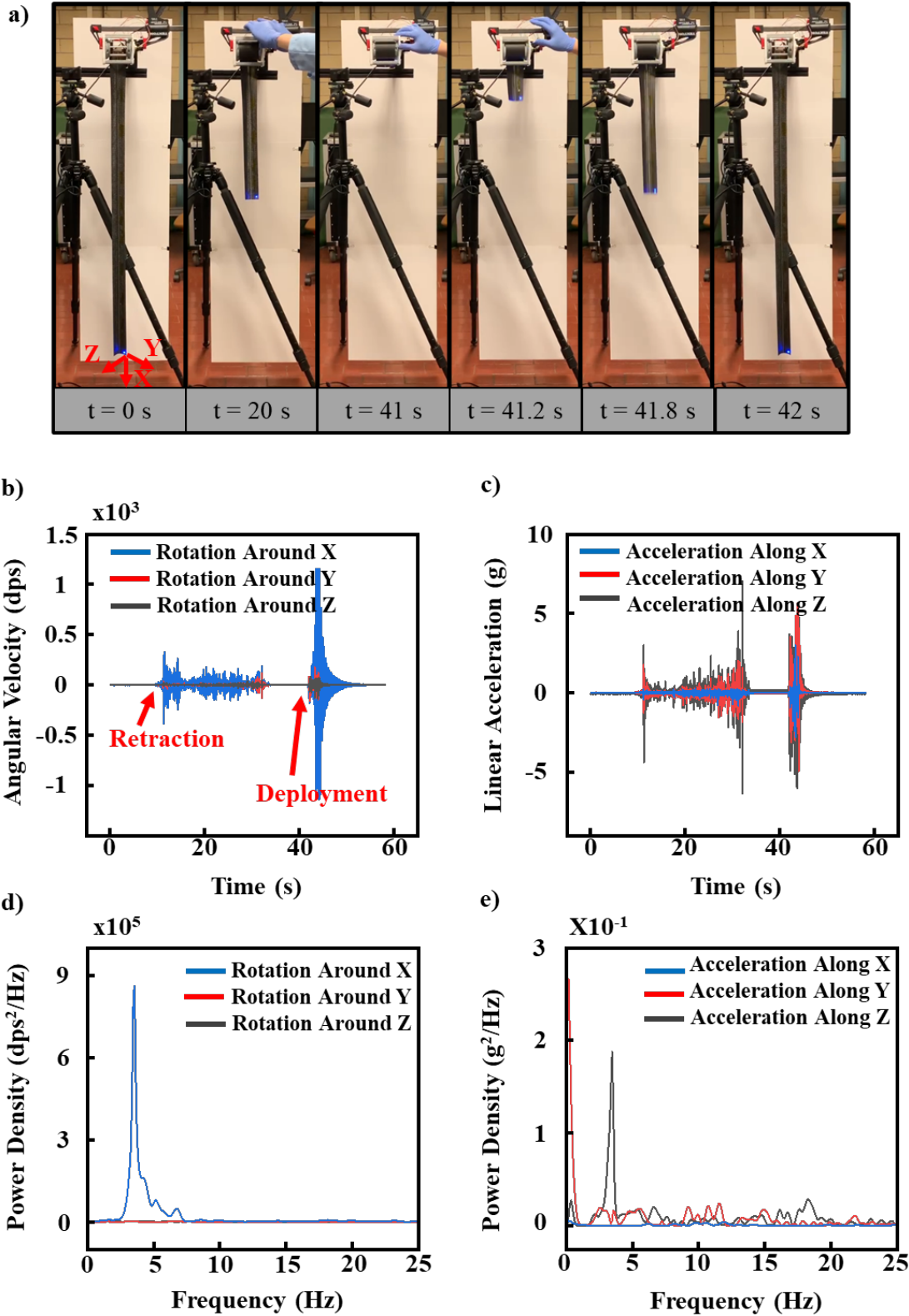

**Fig. 7 Tip-extending Deployment Test and Vibration Analyses.** a) The coiling and self-deployment process. b) and c) Time-domain data of rotation and acceleration during the coiling and self-deployment process. d) and e) Power spectral density plots of the dynamic behavior after full deployment.

### 3.4 Tip-extending Deployment

In the on-orbit deployment test of the boom, the boom will deploy with its root rotating around a spool and its tip extending out to avoid the explosive roll-out deployment. The boom was installed in a deployer as discussed in Section 2.1 and shown in Fig. S1. To better understand the behavior of the boom in the actual scenario, a deployer-assisted, tip-extending deployment was carried out. The deployer included a rotational hub to which the boom root is clamped. The hub had a slip ring to electrically connect the 44-AWG wires to a DAQ system and a power source. The boom was oriented downwards along the gravity vector to minimize gravity-induced deformation and friction. Figure 7a includes the time stamps and images of the retraction and deployment of the boom. Supplemental Video 5 shows the recorded video. It took approximately 0.85 seconds to complete the tip-extending self-deployment, which is 15% shorter than the roll-out deployment.

All gyroscope and accelerometer responses in all three axes were recorded and plotted in Fig. 7b and 7c. PSD results were calculated and plotted for the data points after the deployment (Fig. 7d and 7e) and revealed the $1^{st}$ major natural frequency at 3.56 Hz, which corresponded to the $1^{st}$ twisting mode (significant amplitude of rotation around the $x$-axis). However, the frequency was lower than the previous results shown in Table 2. The decrease in the $1^{st}$ vibration mode frequency was expected because of the change in the clamping boundary condition in the deployer. Compared to the 3D-printed clamp shown in Section 3.2.1, the miniature clamp designed for fitting in the deployer was much smaller in all dimensions, leading to a weaker boundary condition. Additionally, the deployment of the bistable boom caused a shock load at the end of deployment that led to a slight retraction of the boom and inverse rotation of the hub. This made the end-of-deployment vibration complicated compared to the previous tests. Therefore, higher-frequency modes are not presented in the PSD plots.

## 4. Conclusions and Outlook

In this study, we introduced a CubeSat-compatible bistable ultrathin composite boom with integrated flexible electronics for monitoring its deployment dynamics and deployed frequency mode shapes with the overarching goal of demonstrating the concept in orbit. This paper presented the design, fabrication, materials, mechanics, space environmental testing, and on-ground deployment dynamics of a multifunctional bistable boom. The multifunctional bistable booms consisted of ultrathin CFRP shells integrated with flexible electronics on the boom tip and embedded or surface-mount thin wires for power and data transmission. An $I^2C$ communication protocol was successfully set up for data transfer through the thin wires from the boom-tip flexible electronics to the CubeSat DAQ unit through a rotating slip ring. A maximum sampling rate of 208 Hz was supported, which was well above the expected frequency range for the first-order vibrational mode of interest. In addition, the flexible devices showed survivability and stability at various temperatures ranging from 7 °C to 50 °C at a vacuum pressure level of approximately $1 \times 10^{-6}$ Torr. The electrical connection between the boom and the DAQ system through the slip ring was determined to be robust. Ground testing demonstrated the ability of the boom to sense its deployment dynamics and the survivability of the flexible electronics and conductive wires during these aggressive folding and deployment processes. Multiple vibrational modes were detected with ground-based vibration tests. The natural frequencies of the vibration modes captured were consistent among all testing methods: impact-induced vibration, white-noise-background vibration, and linear-frequency-sweep vibration. FEA simulations were used to reveal the physical shape of each vibration mode. Lower order modes were confirmed with the relative vibration amplitude presented in the PSD results ($1^{st}$ twisting mode, $1^{st}$ bending mode, and $2^{nd}$ twisting mode).

A boom unit has been installed on a 3U CubeSat by Virginia Tech that is tentatively scheduled to launch into space circa 2025. During this mission, the boom will be deployed, and the deployment and vibration behavior will be measured using the integrated boom-tip devices. The on-board cameras will record the images of the boom-tip LED during these experiments to confirm the status of the boom. The experiments will provide in-space validation of the functionality of the boom. Since on-ground tests cannot fully simulate the space conditions, the results of on-orbit tests will be compared to the on-ground experiments to gain insights on the effects of space environments such as microgravity and variable thermal conditions on the deployment and dynamics behavior of the boom.

Several future research directions are suggested by the present study. First, although the long-term operation was out of the scope of the flight hardware demonstration, it is important to protect the flexible electronics and the composite booms from space environmental stressors in low-Earth orbits such as atomic oxygen and vacuum ultraviolet radiation. These factors can degrade and ablate the polymer materials. Protective coatings such as silicon dioxides can be used to shield the boom materials from these hostile conditions. Second, it may be desirable to have multiple flexible electronic devices such as the IMUs distributed at different locations of the boom to provide more accurate measurements of the boom deployment status. Multiple IMUs may enable the direct capture of vibrational modes in addition to the frequencies.

## Appendix

## Determining the ABD matrix using the classical laminate theory

The classical laminate theory defines the ABD matrix of the laminate [21]. By definition, the ABD matrix of a laminate is a 6-by-6 stiffness matrix that relates the extensional in-plane strains ($\varepsilon_{xx}, \varepsilon_{yy}, \varepsilon_{xy}$), and the bending curvatures ($\kappa_{xx}, \kappa_{yy}, \kappa_{xy}$) with the applied loads ($N_{xx}, N_{xx}, N_{xx}$), and the applied bending moments ($M_{xx}, M_{xx}, M_{xx}$).

$$\begin{bmatrix} N_{xx} \\ N_{yy} \\ N_{xy} \\ M_{xx} \\ M_{yy} \\ M_{xy} \end{bmatrix} = \begin{bmatrix} A & B \\ B & D \end{bmatrix} \cdot \begin{bmatrix} \varepsilon_{xx} \\ \varepsilon_{yy} \\ \varepsilon_{xy} \\ \kappa_{xx} \\ \kappa_{yy} \\ \kappa_{xy} \end{bmatrix} \qquad (A1)$$

The **A** matrix (3-by-3) represents the extensional and shear stiffness of the laminate by directly relating the in-plain strains with the applied loads. The **D** matrix (3-by-3) represents the bending and torsional stiffness of the laminate by directly relating the bending curvatures with the applied moment. The **B** matrix (3-by-3), on the other hand, represents the coupling effect between extension/shear and bending/torsion. To calculate the **A**, **B**, and **D** matrices, the reduced stiffness matrix, $\boldsymbol{Q_k}$, of the $k^{th}$ ply is first defined:

$$\boldsymbol{Q_k} = \begin{bmatrix} \dfrac{E_{11}^{2}}{E_{11}-\nu_{12}^{2}E_{22}} & \dfrac{\nu_{12}E_{11}E_{22}}{E_{11}-\nu_{12}^{2}E_{22}} & 0 \\ \dfrac{\nu_{12}E_{11}E_{22}}{E_{11}-\nu_{12}^{2}E_{22}} & \dfrac{E_{11}E_{22}}{E_{11}-\nu_{12}^{2}E_{22}} & 0 \\ 0 & 0 & G_{12} \end{bmatrix}, \qquad (A2)$$

where $E_{11}$ is the longitudinal Young's modulus of the $k^{th}$ ply, $E_{22}$ is the transverse Young's modulus of the $k^{th}$ ply, $v_{12}$ is the Poisson's ratio, and $G_{12}$ is the shear modulus. Then the lamina stiffness matrix, $\bar{\boldsymbol{Q}}_k$, is defined as:

$$\bar{\boldsymbol{Q}}_k = \begin{bmatrix} \bar{Q}_{11} & \bar{Q}_{12} & \bar{Q}_{16} \\ \bar{Q}_{12} & \bar{Q}_{22} & \bar{Q}_{26} \\ \bar{Q}_{16} & \bar{Q}_{26} & \bar{Q}_{66} \end{bmatrix}, \tag{A3}$$

where the terms of the $\bar{Q}$ matrix are defined as follow:

$$\begin{aligned}
\bar{Q}_{11} &= Q_{11} \cos^4 \theta + 2(Q_{12} + 2Q_{66}) \cos^2 \theta \sin^2 \theta + Q_{22} \sin^4 \theta \\
\bar{Q}_{12} &= Q_{12}(\cos^4 \theta + \sin^4 \theta) + (Q_{11} + Q_{22} - 4Q_{66}) \cos^2 \theta \sin^2 \theta \\
\bar{Q}_{16} &= (Q_{11} - Q_{12} - 2Q_{66}) \cos^3 \theta \sin \theta - (Q_{22} - Q_{12} - 2Q_{66}) \cos \theta \sin^3 \theta \\
\bar{Q}_{22} &= Q_{11} \sin^4 \theta + 2(Q_{12} + 2Q_{66}) \cos^2 \theta \sin^2 \theta + Q_{22} \cos^4 \theta \\
\bar{Q}_{26} &= (Q_{11} - Q_{12} - 2Q_{66}) \cos \theta \sin^3 \theta - (Q_{22} - Q_{12} - 2Q_{66}) \cos^3 \theta \sin \theta \\
\bar{Q}_{66} &= (Q_{11} + Q_{22} - 2Q_{12} - 2Q_{66}) \cos^2 \theta \sin^2 \theta + Q_{66}(\cos^4 \theta + \sin^4 \theta)
\end{aligned} \tag{A4}$$

where $\theta$ is the fiber orientation of the $k^{th}$ ply.

In addition, it is necessary to define a geometric quantity, $z_k$, to describe the relative position of the $k^{th}$ ply to the mid-plane of the laminate.

$$\begin{aligned} z_0 &= -\frac{t}{2} \\ z_k &= z_{k-1} + t_k, \end{aligned} \tag{A5}$$

where $t$ is the total thickness of the laminate, and $t_k$ is the thickness of the $k^{th}$ ply. The lamina stiffness matrices of each ply are then used to construct the $\boldsymbol{A}$, $\boldsymbol{B}$, and $\boldsymbol{D}$ matrices of the laminate:

$$\begin{aligned}
A_{ij} &= \sum_{k=1}^{n} \{\bar{Q}_{ij}\}_k (z_k - z_{k-1}) \\
B_{ij} &= \frac{1}{2} \sum_{k=1}^{n} \{\bar{Q}_{ij}\}_k (z_k^2 - z_{k-1}^2) \\
D_{ij} &= \frac{1}{3} \sum_{k=1}^{n} \{\bar{Q}_{ij}\}_k (z_k^3 - z_{k-1}^3)
\end{aligned} \tag{A6}$$

In our case, the final values of the $\boldsymbol{A}$, $\boldsymbol{B}$, and $\boldsymbol{D}$ matrices for the laminate without embedded wires are:

$$A = \begin{bmatrix} 11.6440 & 4.9488 & 0 \\ 4.9488 & 6.0654 & 0 \\ 0 & 0 & 5.1166 \end{bmatrix} \text{GPa}$$

$$B = \begin{bmatrix} 0 & 0 & 0 \\ 0 & 0 & 0 \\ 0 & 0 & 0 \end{bmatrix} \text{GPa} \quad (A7)$$

$$D = \begin{bmatrix} 0.0165 & 0.0131 & 0 \\ 0.0131 & 0.0158 & 0 \\ 0 & 0 & 0.0135 \end{bmatrix} \text{GPa}$$

In the case of the zero $B$ matrix, the non-dimensional $\widehat{D}$ matrix reduced to:

$$\widehat{D} = \frac{D}{D_{11}} \quad (A8)$$

## Supplemental Information

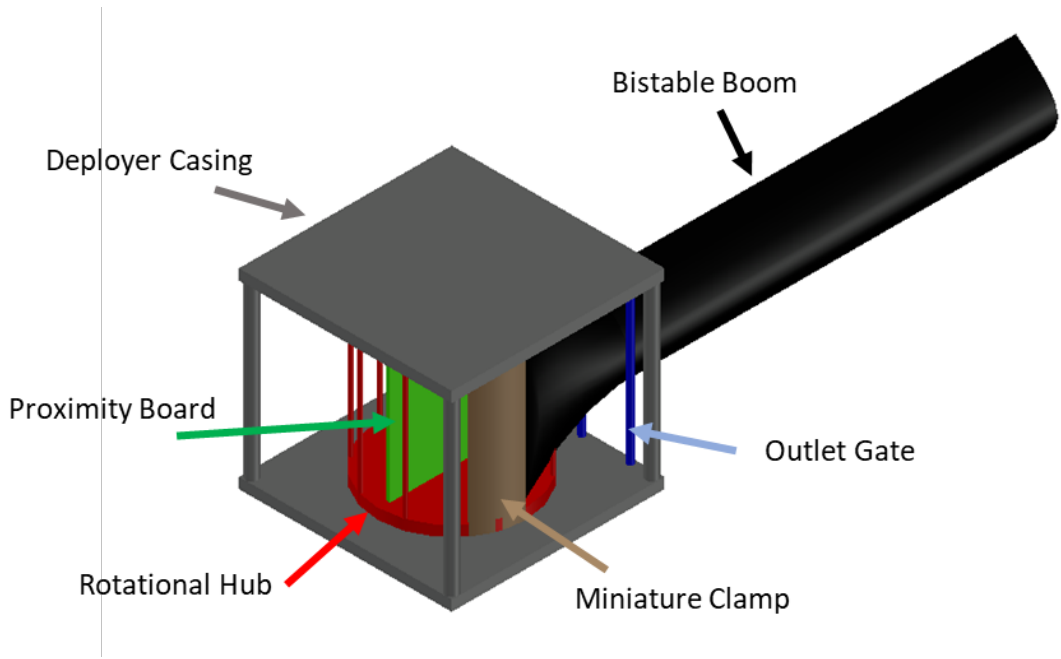

**Fig S1. Schematic of the boom with its root connected to a deployer.**

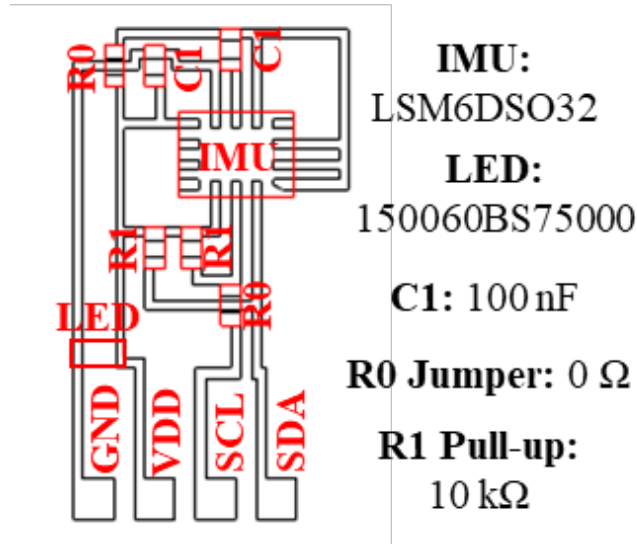

**Fig S2. Circuit design of the boom-tip flexible devices with all electronic component solder positions marked.**

## Acknowledgments

This material is based upon work supported by the Air Force Office of Scientific Research under award number FA9550-22-1-0284 and NASA. Any opinions, findings, and conclusions or recommendations expressed in this material are those of the author(s) and do not necessarily reflect the views of the United States Air Force or NASA. The authors also gratefully acknowledge the support from the Haythornthwaite Foundation Research Initiation Grant and startup funding from the University of Illinois at Urbana-Champaign. We also thank Alexander Ambruso for the help with the TVAC tests.

## Data Availability Statement

The data that support the findings of this study are available from the corresponding author, X.N., upon reasonable request.

## Declaration of Competing Interest

The authors declare that they have no known competing financial interests or personal relationships that could have appeared to influence the work reported in this paper.

## Authors Contributions

Y.Y., X.N., and J.F. conceived the research idea and designed the research. Y.Y. performed the design, fabrication, and integration of the flexible electronic devices, and performed experimental testing, and FEA for the integrated structures. J.F. led the fabrication of the composite structures and the integration between structures and embedded conductive traces. S.B. provided various experimental equipment, supported the thermal–vacuum testing, and provided guidance in electrical testing. Y.Y. analyzed experimental and numerical data. X.N. and Y.Y. wrote and revised the manuscript. J.F. and S.B. revised and commented on the manuscript.